\input harvmac
\input epsf

%
%
%
%
\def\unredoffs{} \def\redoffs{\voffset=-.40truein\hoffset=-.40truein}
\def\speclscape{}
%
%
%
%
\newbox\leftpage \newdimen\fullhsize \newdimen\hstitle \newdimen\hsbody
\tolerance=1000\hfuzz=2pt
\catcode`\@=11 
\def\bigans{b }
\def\answ{b }
\ifx\answ\bigans\message{(This will come out unreduced.}
\magnification=1200\unredoffs\baselineskip=16pt plus 2pt minus 1pt
\hsbody=\hsize \hstitle=\hsize 
\else\message{(This will be reduced.} \let\l@r=L
\magnification=1000\baselineskip=16pt plus 2pt minus 1pt
\vsize=7truein \redoffs
\hstitle=8truein\hsbody=4.75truein\fullhsize=10truein\hsize=\hsbody
\output={\ifnum\pageno=0 
    \shipout\vbox{\speclscape{\hsize\fullhsize\makeheadline}
      \hbox to \fullhsize{\hfill\pagebody\hfill}}\advancepageno
    \else
    \almostshipout{\leftline{\vbox{\pagebody\makefootline}}}\advancepageno
    \fi}
\def\almostshipout#1{\if L\l@r \count1=1 \message{[\the\count0.\the\count1]}
        \global\setbox\leftpage=#1 \global\let\l@r=R
   \else \count1=2
    \shipout\vbox{\speclscape{\hsize\fullhsize\makeheadline}
        \hbox to\fullhsize{\box\leftpage\hfil#1}}  \global\let\l@r=L\fi}
\fi
%
\newcount\yearltd\yearltd=\year\advance\yearltd by -1900

\def\Title#1#2{\nopagenumbers\abstractfont\hsize=\hstitle\rightline{#1}%
\vskip 1in\centerline{\titlefont #2}\abstractfont\vskip
.5in\pageno=0}
%
%

\def\draftmode{\message{ DRAFTMODE }\def\draftdate{{\rm preliminary draft:
\number\month/\number\day/\number\yearltd\ \ \hourmin}}%
\headline={\hfil\draftdate}\writelabels\baselineskip=20pt plus 2pt
minus 2pt
   {\count255=\time\divide\count255 by 60 \xdef\hourmin{\number\count255}
    \multiply\count255 by-60\advance\count255 by\time
    \xdef\hourmin{\hourmin:\ifnum\count255<10 0\fi\the\count255}}}
\def\nolabels{\def\wrlabeL##1{}\def\eqlabeL##1{}\def\reflabeL##1{}}
\def\writelabels{\def\wrlabeL##1{\leavevmode\vadjust{\rlap{\smash%
{\line{{\escapechar=` \hfill\rlap{\sevenrm\hskip.03in\string##1}}}}}}}%
\def\eqlabeL##1{{\escapechar-1\rlap{\sevenrm\hskip.05in\string##1}}}%
\def\reflabeL##1{\noexpand\llap{\noexpand\sevenrm\string\string\string##1}}}
\nolabels
%
\global\newcount\secno \global\secno=0 \global\newcount\meqno
\global\meqno=1
\def\newsec#1{\global\advance\secno by1\message{(\the\secno. #1)}
\global\subsecno=0\eqnres@t\noindent{\bf\the\secno. #1}
\writetoca{{\secsym} {#1}}\par\nobreak\medskip\nobreak}
\def\eqnres@t{\xdef\secsym{\the\secno.}\global\meqno=1\bigbreak\bigskip}
\def\sequentialequations{\def\eqnres@t{\bigbreak}}\xdef\secsym{}
\global\newcount\subsecno \global\subsecno=0
\def\subsec#1{\global\advance\subsecno by1\message{(\secsym\the\subsecno. #1)}
\ifnum\lastpenalty>9000\else\bigbreak\fi
\noindent{\it\secsym\the\subsecno. #1}\writetoca{\string\quad
{\secsym\the\subsecno.} {#1}}\par\nobreak\medskip\nobreak}
\def\appendix#1#2{\global\meqno=1\global\subsecno=0\xdef\secsym{\hbox{#1.}}
\bigbreak\bigskip\noindent{\bf Appendix #1. #2}\message{(#1. #2)}
\writetoca{Appendix {#1.} {#2}}\par\nobreak\medskip\nobreak}
%
%
\def\eqnn#1{\xdef #1{(\secsym\the\meqno)}\writedef{#1\leftbracket#1}%
\global\advance\meqno by1\wrlabeL#1}
\def\eqna#1{\xdef #1##1{\hbox{$(\secsym\the\meqno##1)$}}
\writedef{#1\numbersign1\leftbracket#1{\numbersign1}}%
\global\advance\meqno by1\wrlabeL{#1$\{\}$}}
\def\eqn#1#2{\xdef #1{(\secsym\the\meqno)}\writedef{#1\leftbracket#1}%
\global\advance\meqno by1$$#2\eqno#1\eqlabeL#1$$}
%
\newskip\footskip\footskip14pt plus 1pt minus 1pt 
\def\footnotefont{\ninepoint}\def\f@t#1{\footnotefont #1\@foot}
\def\f@@t{\baselineskip\footskip\bgroup\footnotefont\aftergroup\@foot\let\next}
\setbox\strutbox=\hbox{\vrule height9.5pt depth4.5pt width0pt}
\global\newcount\ftno \global\ftno=0
\def\foot{\global\advance\ftno by1\footnote{$^{\the\ftno}$}}
%
\newwrite\ftfile
\def\footend{\def\foot{\global\advance\ftno by1\chardef\wfile=\ftfile
$^{\the\ftno}$\ifnum\ftno=1\immediate\openout\ftfile=foots.tmp\fi%
\immediate\write\ftfile{\noexpand\smallskip%
\noexpand\item{f\the\ftno:\ }\pctsign}\findarg}%
\def\footatend{\vfill\eject\immediate\closeout\ftfile{\parindent=20pt
\centerline{\bf Footnotes}\nobreak\bigskip\input foots.tmp }}}
\def\footatend{}
%
%
\global\newcount\refno \global\refno=1
\newwrite\rfile
\def\ref{[\the\refno]\nref}
\def\nref#1{\xdef#1{[\the\refno]}\writedef{#1\leftbracket#1}%
\ifnum\refno=1\immediate\openout\rfile=refs.tmp\fi
\global\advance\refno by1\chardef\wfile=\rfile\immediate
\write\rfile{\noexpand\item{#1\
}\reflabeL{#1\hskip.31in}\pctsign}\findarg}
\def\findarg#1#{\begingroup\obeylines\newlinechar=`\^^M\pass@rg}
{\obeylines\gdef\pass@rg#1{\writ@line\relax #1^^M\hbox{}^^M}%
\gdef\writ@line#1^^M{\expandafter\toks0\expandafter{\striprel@x #1}%
\edef\next{\the\toks0}\ifx\next\em@rk\let\next=\endgroup\else\ifx\next\empty%
\else\immediate\write\wfile{\the\toks0}\fi\let\next=\writ@line\fi\next\relax}}
\def\striprel@x#1{} \def\em@rk{\hbox{}}
\def\lref{\begingroup\obeylines\lr@f}
\def\lr@f#1#2{\gdef#1{\ref#1{#2}}\endgroup\unskip}

\def\addref#1{\immediate\write\rfile{\noexpand\item{}#1}} 
\def\startrefs#1{\immediate\openout\rfile=refs.tmp\refno=#1}
\def\xref{\expandafter\xr@f}\def\xr@f[#1]{#1}
\def\refs#1{\count255=1[\r@fs #1{\hbox{}}]}
\def\r@fs#1{\ifx\und@fined#1\message{reflabel \string#1 is undefined.}%
\nref#1{need to supply reference \string#1.}\fi%
\vphantom{\hphantom{#1}}\edef\next{#1}\ifx\next\em@rk\def\next{}%
\else\ifx\next#1\ifodd\count255\relax\xref#1\count255=0\fi%
\else#1\count255=1\fi\let\next=\r@fs\fi\next}
%

%
\newwrite\ffile\global\newcount\figno \global\figno=1
\def\fig{fig.~\the\figno\nfig}
\def\nfig#1{\xdef#1{fig.~\the\figno}%
\writedef{#1\leftbracket fig.\noexpand~\the\figno}%
\ifnum\figno=1\immediate\openout\ffile=figs.tmp\fi\chardef\wfile=\ffile%
\immediate\write\ffile{\noexpand\medskip\noexpand\item{Fig.\
\the\figno. }
\reflabeL{#1\hskip.55in}\pctsign}\global\advance\figno
by1\findarg}
\def\xfig{\expandafter\xf@g}\def\xf@g fig.\penalty\@M\ {}
\def\figs#1{figs.~\f@gs #1{\hbox{}}}
\def\f@gs#1{\edef\next{#1}\ifx\next\em@rk\def\next{}\else
\ifx\next#1\xfig #1\else#1\fi\let\next=\f@gs\fi\next}
\newwrite\lfile
{\escapechar-1\xdef\pctsign{\string\%}\xdef\leftbracket{\string\{}
\xdef\rightbracket{\string\}}\xdef\numbersign{\string\#}}

\def\writestop{\def\writestoppt{\immediate\write\lfile{\string\pageno%
\the\pageno\string\startrefs\leftbracket\the\refno\rightbracket%
\string\def\string\secsym\leftbracket\secsym\rightbracket%
\string\secno\the\secno\string\meqno\the\meqno}\immediate\closeout\lfile}}
\def\writestoppt{}\def\writedef#1{}
\def\seclab#1{\xdef #1{\the\secno}\writedef{#1\leftbracket#1}\wrlabeL{#1=#1}}
\def\subseclab#1{\xdef #1{\secsym\the\subsecno}%
\writedef{#1\leftbracket#1}\wrlabeL{#1=#1}}
\newwrite\tfile \def\writetoca#1{}
\def\leaderfill{\leaders\hbox to 1em{\hss.\hss}\hfill}
\def\writetoc{\immediate\openout\tfile=toc.tmp
     \def\writetoca##1{{\edef\next{\write\tfile{\noindent ##1
     \string\leaderfill {\noexpand\number\pageno} \par}}\next}}}

%
\catcode`\@=12 
%
\edef\tfontsize{\ifx\answ\bigans scaled\magstep3\else
scaled\magstep4\fi} \font\titlerm=cmr10 \tfontsize
\font\titlerms=cmr7 \tfontsize \font\titlermss=cmr5 \tfontsize
\font\titlei=cmmi10 \tfontsize \font\titleis=cmmi7 \tfontsize
\font\titleiss=cmmi5 \tfontsize \font\titlesy=cmsy10 \tfontsize
\font\titlesys=cmsy7 \tfontsize \font\titlesyss=cmsy5 \tfontsize
\font\titleit=cmti10 \tfontsize \skewchar\titlei='177
\skewchar\titleis='177 \skewchar\titleiss='177
\skewchar\titlesy='60 \skewchar\titlesys='60
\skewchar\titlesyss='60
\def\titlefont{\def\rm{\fam0\titlerm}
\textfont0=\titlerm \scriptfont0=\titlerms
\scriptscriptfont0=\titlermss \textfont1=\titlei
\scriptfont1=\titleis \scriptscriptfont1=\titleiss
\textfont2=\titlesy \scriptfont2=\titlesys
\scriptscriptfont2=\titlesyss \textfont\itfam=\titleit
\def\it{\fam\itfam\titleit}\rm}
 \ifx\answ\bigans\else scaled\magstep1\fi
\ifx\answ\bigans\def\abstractfont{\tenpoint}\else
\font\abssl=cmsl10 scaled \magstep1 \font\absrm=cmr10
scaled\magstep1 \font\absrms=cmr7 scaled\magstep1
\font\absrmss=cmr5 scaled\magstep1 \font\absi=cmmi10
scaled\magstep1 \font\absis=cmmi7 scaled\magstep1
\font\absiss=cmmi5 scaled\magstep1 \font\abssy=cmsy10
scaled\magstep1 \font\abssys=cmsy7 scaled\magstep1
\font\abssyss=cmsy5 scaled\magstep1 \font\absbf=cmbx10
scaled\magstep1 \skewchar\absi='177 \skewchar\absis='177
\skewchar\absiss='177 \skewchar\abssy='60 \skewchar\abssys='60
\skewchar\abssyss='60
\def\abstractfont{\def\rm{\fam0\absrm}
\textfont0=\absrm \scriptfont0=\absrms \scriptscriptfont0=\absrmss
\textfont1=\absi \scriptfont1=\absis \scriptscriptfont1=\absiss
\textfont2=\abssy \scriptfont2=\abssys \scriptscriptfont2=\abssyss
\textfont\itfam=\bigit \def\it{\fam\itfam\bigit}\def\footnotefont{\tenpoint}%
\textfont\slfam=\abssl \def\sl{\fam\slfam\abssl}%
\textfont\bffam=\absbf \def\bf{\fam\bffam\absbf}\rm}\fi
\def\tenpoint{\def\rm{\fam0\tenrm}
\textfont0=\tenrm \scriptfont0=\sevenrm \scriptscriptfont0=\fiverm
\textfont1=\teni  \scriptfont1=\seveni  \scriptscriptfont1=\fivei
\textfont2=\tensy \scriptfont2=\sevensy \scriptscriptfont2=\fivesy
\textfont\itfam=\tenit \def\it{\fam\itfam\tenit}\def\footnotefont{\ninepoint}%
\textfont\bffam=\tenbf
\def\bf{\fam\bffam\tenbf}\def\sl{\fam\slfam\tensl}\rm}
\font\ninerm=cmr9 \font\sixrm=cmr6 \font\ninei=cmmi9
\font\sixi=cmmi6 \font\ninesy=cmsy9 \font\sixsy=cmsy6
\font\ninebf=cmbx9 \font\nineit=cmti9 \font\ninesl=cmsl9
\skewchar\ninei='177 \skewchar\sixi='177 \skewchar\ninesy='60
\skewchar\sixsy='60
\def\ninepoint{\def\rm{\fam0\ninerm}
\textfont0=\ninerm \scriptfont0=\sixrm \scriptscriptfont0=\fiverm
\textfont1=\ninei \scriptfont1=\sixi \scriptscriptfont1=\fivei
\textfont2=\ninesy \scriptfont2=\sixsy \scriptscriptfont2=\fivesy
\textfont\itfam=\ninei \def\it{\fam\itfam\nineit}\def\sl{\fam\slfam\ninesl}%
\textfont\bffam=\ninebf \def\bf{\fam\bffam\ninebf}\rm}
%
%

\hyphenation{anom-aly anom-alies coun-ter-term coun-ter-terms}
\def\inv{^{\raise.15ex\hbox{${\scriptscriptstyle -}$}\kern-.05em 1}}

\def\Dsl{\,\raise.15ex\hbox{/}\mkern-13.5mu D} 
\def\dsl{\raise.15ex\hbox{/}\kern-.57em\partial}

\font\bigit=cmti10 scaled \magstep1
\def\lspace{\ifx\answ\bigans{}\else\qquad\fi}
\def\lbspace{\ifx\answ\bigans{}\else\hskip-.2in\fi} 
\def\boxeqn#1{\vcenter{\vbox{\hrule\hbox{\vrule\kern3pt\vbox{\kern3pt
      \hbox{${\displaystyle #1}$}\kern3pt}\kern3pt\vrule}\hrule}}}
\def\mbox#1#2{\vcenter{\hrule \hbox{\vrule height#2in
          \kern#1in \vrule} \hrule}}  
%

\def\darr#1{\raise1.5ex\hbox{$\leftrightarrow$}\mkern-16.5mu #1}

\def\roughly#1{\raise.3ex\hbox{$#1$\kern-.75em\lower1ex\hbox{$\sim$}}}

\def\IB{\relax\hbox{$\inbar\kern-.3em{\rm B}$}}

\def\ID{\relax\hbox{$\inbar\kern-.3em{\rm D}$}}
\def\IE{\relax\hbox{$\inbar\kern-.3em{\rm E}$}}
\def\IF{\relax\hbox{$\inbar\kern-.3em{\rm F}$}}
\def\IG{\relax\hbox{$\inbar\kern-.3em{\rm G}$}}
\def\IGa{\relax\hbox{${\rm I}\kern-.18em\Gamma$}}
\def\IH{\relax{\rm I\kern-.18em H}}
\def\IK{\relax{\rm I\kern-.18em K}}
\def\IL{\relax{\rm I\kern-.18em L}}
\def\IP{\relax{\rm I\kern-.18em P}}
\def\IR{{\bf R}}

\def\II{\relax{\rm I\kern-.18em I}}






\def\inbar{\,\vrule height1.5ex width.4pt depth0pt}


\def\ve{{\varepsilon}}

\def\lref{\begingroup\obeylines\lr@f}
\def\lr@f#1#2{\gdef#1{\ref#1{#2}}\endgroup\unskip}

\lref\gsw{M.~Green, J.~Schwarz, E.~Witten, ``Superstring theory'',
 Cambridge Univeristy Press, 1987~.}

\lref\bdl{
M.~Berkooz, M.~R.~Douglas and R.~G.~Leigh,
``Branes intersecting at angles,''
Nucl.\ Phys.\ B {\bf 480}, 265 (1996)
[arXiv:hep-th/9606139].
}

\lref\HaSe{
A.~Hashimoto and S.~Sethi,
``Holography and string dynamics in time-dependent backgrounds,''
arXiv:hep-th/0208126.
}

\lref\tw{E.F.~Taylor and J.A.~Wheeler, ``Spacetime Physics,''
2nd edition, Freeman, 1992~.}

\lref\hs{
G.~T.~Horowitz and A.~R.~Steif,
``Singular String Solutions With Nonsingular Initial Data,''
Phys.\ Lett.\ B {\bf 258}, 91 (1991).
}

\lref\Khoury{
J.~Khoury, B.~A.~Ovrut, N.~Seiberg, P.~J.~Steinhardt and N.~Turok,
``From big crunch to big bang,''
Phys.\ Rev.\ D {\bf 65}, 086007 (2002)
[arXiv:hep-th/0108187].
}

\lref\lmsI{ H.~Liu, G.~Moore and N.~Seiberg, ``Strings in a
time-dependent orbifold,'' arXiv:hep-th/0204168.}

\lref\lmsII{H.~Liu, G.~Moore and N.~Seiberg, ``Strings in
Time-Dependent Orbifolds,'' arXiv:hep-th/0206182.}

\lref\Ff{
J.~M.~Figueroa-O'Farrill,
``Breaking the M-waves,''
Class.\ Quant.\ Grav.\  {\bf 17}, 2925 (2000)
[arXiv:hep-th/9904124].
}

\lref\FF{
J.~Figueroa-O'Farrill and J.~Simon,
``Generalized supersymmetric fluxbranes,''
JHEP {\bf 0112}, 011 (2001)
[arXiv:hep-th/0110170].
}

\lref\Cornalba{
L.~Cornalba and M.~S.~Costa,
``A New Cosmological Scenario in String Theory,''
arXiv:hep-th/0203031.
}

\lref\Nekrasov{
N.~Nekrasov,
``Milne universe, tachyons, and quantum group,''
arXiv:hep-th/0203112.
}

\lref\Simon{
J.~Simon,
``The geometry of null rotation identifications,''
JHEP {\bf 0206}, 001 (2002)
[arXiv:hep-th/0203201].
}

\lref\HoPo{ G.~T.~Horowitz and J.~Polchinski, ``Instability of
Spacelike and Null Orbifold Singularities,'' arXiv:hep-th/0206228.
}

\lref\Kou{
L.~Cornalba, M.~S.~Costa and C.~Kounnas,
``A resolution of the cosmological singularity with orientifolds,''
Nucl.\ Phys.\ B {\bf 637}, 378 (2002)
[arXiv:hep-th/0204261].
}

\lref\Craps{
B.~Craps, D.~Kutasov and G.~Rajesh,
``String propagation in the presence of cosmological singularities,''
JHEP {\bf 0206}, 053 (2002)
[arXiv:hep-th/0205101].
}

\lref\Elitzur{
S.~Elitzur, A.~Giveon, D.~Kutasov and E.~Rabinovici,
``From big bang to big crunch and beyond,''
JHEP {\bf 0206}, 017 (2002)
[arXiv:hep-th/0204189].
}

\lref\Law{
A.~Lawrence,
``On the instability of 3D null singularities,''
arXiv:hep-th/0205288.
}

\lref\Fabinger{
M.~Fabinger and J.~McGreevy,
``On Smooth Time-Dependent Orbifolds and Null Singularities,''
arXiv:hep-th/0206196.
}

\lref\gh{
G.~W.~Gibbons and C.~A.~Herdeiro,
``Born-Infeld theory and stringy causality,''
Phys.\ Rev.\ D {\bf 63}, 064006 (2001)
[arXiv:hep-th/0008052].
}

\lref\Abou{
A.~Abouelsaood, C.~G.~Callan, C.~R.~Nappi and S.~A.~Yost,
``Open Strings In Background Gauge Fields,''
Nucl.\ Phys.\ B {\bf 280}, 599 (1987).
}

\lref\rev{C.~P.~Bachas,
``Lectures on D-branes,''
arXiv:hep-th/9806199.
}

\lref\dyn{C.~Bachas,
``D-brane dynamics,''
Phys.\ Lett.\ B {\bf 374}, 37 (1996)
[arXiv:hep-th/9511043].
}

\lref\bp{C.~Bachas and M.~Porrati,
``Pair Creation Of Open Strings In An Electric Field,''
Phys.\ Lett.\ B {\bf 296}, 77 (1992)
[arXiv:hep-th/9209032];
C.~Bachas, ``Schwinger effect in string theory,''
arXiv:hep-th/9303063.
}

\lref\Acha{
B.~S.~Acharya, J.~M.~Figueroa-O'Farrill, B.~Spence and S.~Stanciu,
``Planes, branes and automorphisms. II: Branes in motion,''
JHEP {\bf 9807}, 005 (1998)
[arXiv:hep-th/9805176].
}

\lref\MolinsXE{
J.~Molins and J.~Simon,
``BPS states and automorphisms,''
Phys.\ Rev.\ D {\bf 62}, 125019 (2000)
[arXiv:hep-th/0007253].
}

\lref\Figue{
J.~M.~Figueroa-O'Farrill,
``Intersecting brane geometries,''
J.\ Geom.\ Phys.\  {\bf 35}, 99 (2000)
[arXiv:hep-th/9806040].
}

\lref\bn{ C. Bachas and N. Nekrasov, may  appear~.}

\lref\Cal{
C.~G.~Callan, J.~M.~Maldacena and A.~W.~Peet,
``Extremal Black Holes As Fundamental Strings,''
Nucl.\ Phys.\ B {\bf 475}, 645 (1996)
[arXiv:hep-th/9510134].
}

\lref\Dab{
A.~Dabholkar, J.~P.~Gauntlett, J.~A.~Harvey and D.~Waldram,
``Strings as Solitons and  Black Holes as Strings,''
Nucl.\ Phys.\ B {\bf 474}, 85 (1996)
[arXiv:hep-th/9511053].
}

\lref\Ma{
D.~Mateos and P.~K.~Townsend,
``Supertubes,''
Phys.\ Rev.\ Lett.\  {\bf 87}, 011602 (2001)
[arXiv:hep-th/0103030].
}

\lref\Cho{
J.~H.~Cho and P.~Oh,
``Super D-helix,''
Phys.\ Rev.\ D {\bf 64}, 106010 (2001)
[arXiv:hep-th/0105095].
}

\lref\Lun{
O.~Lunin and S.~D.~Mathur,
``Metric of the multiply wound rotating string,''
Nucl.\ Phys.\ B {\bf 610}, 49 (2001)
[arXiv:hep-th/0105136].
}

\lref\Mat{
D.~Mateos, S.~Ng and P.~K.~Townsend,
``Supercurves,''
Phys.\ Lett.\ B {\bf 538}, 366 (2002)
[arXiv:hep-th/0204062].
}

\lref\Dav{
J.~R.~David, G.~Mandal and S.~R.~Wadia,
``Microscopic formulation of black holes in string theory,''
Phys.\ Rept.\  {\bf 369}, 549 (2002)
[arXiv:hep-th/0203048].
}

\lref\Dudas{
E.~Dudas, J.~Mourad and C.~Timirgaziu,
``Time and space dependent backgrounds from nonsupersymmetric strings,''
arXiv:hep-th/0209176.
}

\lref\ve{
G.~Veneziano,
``String cosmology: The pre-big bang scenario,''
arXiv:hep-th/0002094.
}

\lref\ek{
J.~Khoury, B.~A.~Ovrut, P.~J.~Steinhardt and N.~Turok,
``The ekpyrotic universe: Colliding branes and the origin
 of the hot big  bang,''
Phys.\ Rev.\ D {\bf 64}, 123522 (2001)
[arXiv:hep-th/0103239].
}


\def\ens{{\it LPTENS,
24 rue Lhomond, 75231 Paris cedex 05, France}}
\def\qmc{{\it Queen Mary, University of London, Mile End Road, London E1 4NS,
UK}}


\Title{\vbox{\baselineskip 10pt  \hbox{LPTENS-02/53}\smallskip
\hbox{QMUL-PH-02-17}\smallskip
 \hbox{hep-th/0210269} {\hbox{ }}}} {\vbox{\vskip -30
true pt \centerline{Null Brane Intersections}
\smallskip
\smallskip
\smallskip
   \smallskip\smallskip
\medskip
\vskip4pt }} \vskip -20 true pt \centerline{Constantin  Bachas$^{\; \dagger}$ and
Chris Hull$^{\; *}$}
\smallskip\smallskip
\centerline{$^{\dagger}$\ens \foot{Unit{\'e} mixte de Recherche du CNRS 
et de l' Ecole Normale Sup{\'e}rieure.}
} \centerline{$^{*}$\qmc}
\medskip \centerline{\tt e-mails: bachas@physique.ens.fr, C.M.Hull@qmul.ac.uk}
\bigskip
\bigskip

{\bf Abstract}: We study pairs of planar D-branes intersecting on  null
hypersurfaces, and  other related  configurations.
These are supersymmetric and have finite energy density. 
They provide open-string analogues of the parabolic orbifold and 
of the null-fluxbrane backgrounds for closed superstrings. 
We derive the spectrum of  open strings, showing in particular
 that if the D-branes are shifted in a spectator dimension  so that they do 
not intersect,  the open strings joining them  have no asymptotic 
states. As a result,  a single non-BPS excitation can in this case
catalyze a condensation of massless  modes, changing significantly the 
underlying supersymmetric vacuum state.  
We argue that a similar phenomenon
can modify  the null cosmological singularity of  the 
time-dependent orbifolds. 
This is a stringy mechanism, distinct from black-hole formation and other 
strong gravitational  instabilities, and  one that
 should  dominate  at weak string coupling. 
A by-product of our analysis is  a new understanding of  the
appearance of 1/4 BPS threshold bound states, 
 at special points in the moduli
space of toroidally-compactified type-II string theory.


   \vfil\break 

\newsec{Introduction}

The problems of time dependence and cosmology in string theory have
recently received considerable attention. A deceptively simple class of
time-dependent backgrounds are orbifolds involving boosts or 
null boosts, rather than spatial  rotations
 \refs {\hs\Khoury\FF\Cornalba\Nekrasov\Simon\lmsI
\Elitzur\Kou\Craps\Law\lmsII\Fabinger\HoPo\HaSe {--} \Dudas }.
These are toy models for  a cosmological bounce, which is a neccessary
ingredient of the pre-Big-Bang \ve\ and  ekpyrotic \ek\ scenarios. 
As has been argued  however in references \refs {\lmsI,\Law,\lmsII,\HoPo} , 
strong gravitational effects raise  serious questions as to
  the validity of 
a perturbative analysis of the spacelike or null singularities in   
these backgrounds. The fate of such
singularities in string theory remains at present  an open problem.  

  A good starting point for addressing these issues
 \refs {\Simon,\lmsI}  is  the parabolic orbifold \hs\ and the
associated null flux brane \FF . These closed-string backgrounds
are supersymmetric, and  have no closed time-like curves. They thus 
avoid many of the obvious difficulties present in  other time-dependent
backgrounds.  The geometries of interest are  orbifolds of 
flat Minkowski space $\IR^{9,1}$ under the action of a Poincar{\'e}
isometry $\Lambda{\cal T}$ , consisting of a
Lorentz transformation  $\Lambda\in SO(9,1)$, 
combined with a translation  ${\cal T}$ that commutes with $\Lambda$.
The null flux brane is obtained when   $\Lambda$ is a 
null boost in $SO(2,1)$   and ${\cal T}$  a translation by
$2\pi r$ in  the remaining $\IR^ 7$. 
The parabolic orbifold is the special case  $r=0$. 
The null flux brane is a  non-singular 
geometry, which does not suffer from the strong-coupling instabilities 
of \refs{\lmsII,\HoPo}  for given  $r$ and sufficiently weak  string
coupling. To a  fundamental string probe, on the other hand, a 
flux brane with $r\sim\sqrt{\alpha^\prime}$ should look
indistinguishable from the  parabolic orbifold background.
Thus there is legitimate hope that one may
understand  the nature of the null singularity within string
perturbation theory in this simple model.

  Motivated by such questions, we  analyze in the present work an
open-string analogue of the parabolic orbifold and null flux brane
backgrounds. The system consists of two planar D-branes of type-II
string theory,  whose worldvolumes are related by the Poincar{\'e}
transformation 
 $\Lambda  {\cal T}$.  The two branes are  in relative motion
and make  a non-zero angle with each other, and  the
minimal distance between them is $2\pi r \equiv b$. When $b=0$
they intersect on a null hypersurface, which  plays  a similar  role 
to  the null singularity in  the parabolic orbifold. 
A simple  example of such a background
has two linear D-strings oriented and
moving so that the point at which  they intersect
 propagates at the speed of light. 
We call this the null-scissors configuration, 
and its generalisation to non-zero $b$  will
be referred to as the shifted null
scissors.
Such configurations preserve 1/2 of the supersymmetries of the individual
branes,\foot{This is  shown in reference \Acha\ , which contains a
general analysis of unbroken supersymmetries for  pairs  of branes related by
an arbitrary element of the Lorentz group $SO(9,1)$.
}
for the same algebraic
reason that determines that
 the parabolic orbifold and flux brane preserve 1/2 of the 32
supersymmetries of flat spacetime. Furthermore, open strings with one
end on each D-brane are directly analogous  to the twisted closed  strings 
of  \refs{\lmsI,\Fabinger}  in the 
corresponding orbifold backgrounds. Open strings with both ends
on the same D-brane correspond likewise to untwisted closed strings. 

  One of the main messages of our work is that twisted closed strings and
  their open-string counterparts will trigger an instability of the
  underlying vacuum, if  produced during  a collision process. The reason,
  as we will show, is that these  strings have no normalizable
  asymptotic  states unless
  $r=0$ in  the orbifold, or $b=0$ for the D-branes.
 In the shifted null scissors
  configuration, this mechanism will force the D-strings to intersect,
  and then recombine as in a standard string interaction. 
We expect similar phenomena  in the null flux brane background, 
  for  $r\sim \sqrt{\alpha^\prime}$ and
  sufficiently weak string coupling constant. In this regime
  there will be  no formation of  black holes  \HoPo\  , and the above
  stringy mechanism could dominate. Ultimately, one would of course
like to have a good   effective description of the 
null D-brane intersection, and  of the related null
Big-Crunch/Big-Bang singularity,  at weak string coupling. It could be
that such a description must be of a statistical  nature. 
Our  analysis  in this paper is aimed at  identifying  the relevant modes,
and should be considered as a step in this direction.

 The plan of this paper is as follows~: in section 2 we introduce the
 basic null-scissors and shifted null-scissors configurations of type-II string theory, 
 and compare them  to the more
 familiar cases of static branes at angles, and of parallel but moving branes.     
 In section 3 we describe various dual configurations, and 
 explain why  they are  1/4 supersymmetric.   We discuss in  detail
 a particular T-dual configuration, consisting of  a pair of static parallel
 D3-branes, one of which carries an  infinite-wavelength plane electromagnetic wave on
 its worldvolume.  The analogs of twisted (untwisted)  closed
 strings are open strings that are charged (neutral)  with respect to
 the electromagnetic field.  In section 4 we analyze  the
 open-string excitations on the  D-branes. Adapting the discussion of
 reference  \lmsI ,  we show that the only normalizable
 states of  the charged open strings have their momenta  pointing in
 the direction of the electromagnetic wave and  are
  massless. Such states thus only exist  for zero shift, 
  or for $r=0$ in the closed-string background.  
 In section 5, we wrap the null-intersecting D-strings on a torus, which is
possible only for special values of the torus moduli, and
show that the two D-strings can  form a supersymmetric bound state at
threshold. This sheds new light on  the appearance of 1/4 BPS threshold bound
states at special points in the moduli space of toroidally compactified type
II strings. Finally, in section 6 we turn to the important  issue of stability.
We first discuss supersymmetric 
generalizations  of the null D-brane scissors, in which
the D-strings carry arbitrary waves,  all travelling in the same
direction as the intersection point. We then explain how a low-energy
collision process can trigger a condensation of massless modes,
leading to a modification of the vacuum that corresponds to
 the splitting and joining of the two D-strings.  We comment on the
analogous orbifold problem and suggest directions for future work.


\newsec{Null Scissors}

 Consider the configuration of Figure 1. An infinite straight rigid rod
makes an angle  $\theta$ with a `reference'   rod, and moves with
uniform velocity $v$ in the normal (downward) direction. The
reference rod is at rest and extends  along  the $x^1$ axis, while
the other moves in the ($x^1$, $x^2$) plane. As can be verified
easily, the intersection point (I)   of these  two rods propagates
with velocity $v_{\rm I} = v/{\rm sin}\theta$ in the (negative)
$x^1$ direction. There are thus  three inequivalent possibilities:
for $v$ greater, less,  or equal to $c\;{\rm sin}\theta$, the
intersection velocity  is greater, less,  or equal to the speed of
light.

\bigskip
\ifig\moverot{A D1-brane,   rotated by an angle
 $\theta$ relative to the $x^1$ axis,
and moving in the normal direction with uniform velocity $v$. Its
intersection with a static reference D1-brane propagates along the
negative $x^1$ direction with  speed $v_{\rm I}=v/{\rm sin}\theta$.
} {\epsfxsize3.1in\epsfbox{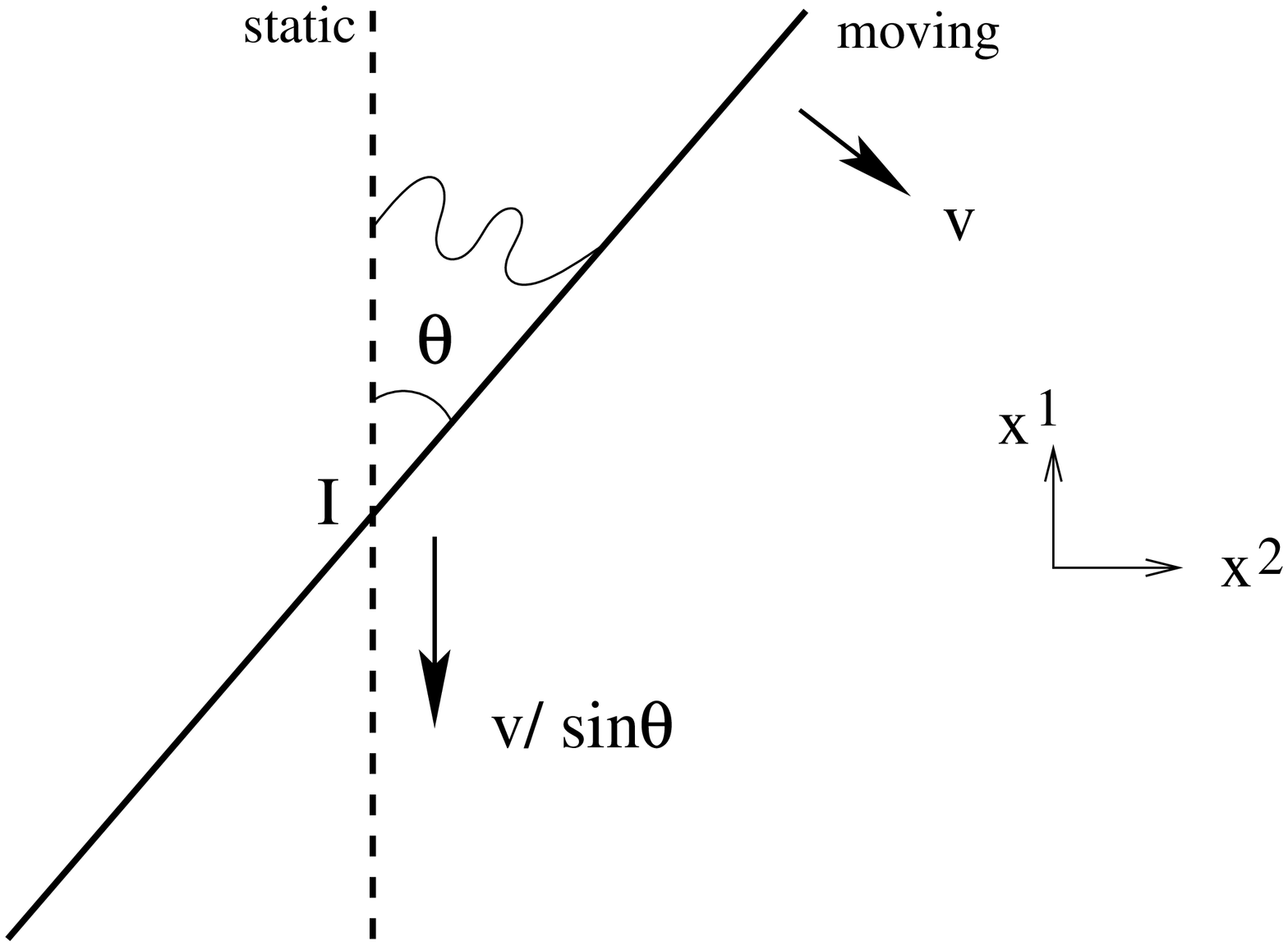}}


   For a given $v$ and  sufficiently small angle $\theta$,
we have $v_{\rm I}>c$~ (henceforth we set $c=1$)~. In   books on
special relativity this is sometimes called  the
scissors\foot{This is a slight misnomer because in common scissors
the intersection point does not move.} paradox \tw . There isn't
of course  anything paradoxical in this example: the
superluminal propagation of the intersection point  cannot be used
to send a signal faster than light. One could for instance attach
one end of a string to one of the   rods and the other end to the other rod  
and let the string move down with the
 intersection point. However, rather
than keep pace  with this
 latter,  the string  will either break or  bend the
rods, and slow
 or stop their relative motion.
 
 The above configuration can be realized in type IIB string theory,
where the  rods may  be infinite straight D1-branes. We assume for
the moment that both D1-branes sit at $x^3=\cdots = x^9=0$. In the
subspace spanned by $x^0, x^1, x^2$ their worldvolume embeddings
are described by the following equations:
$${\underline{\rm static}:}\ \ x^2=0\ , $$
\eqn\moving{ {\underline{\rm moving}:}\ \ x^2 = (v\; x^0 + {\rm
sin}\theta\; x^1)/ {\rm cos}\theta \ ,  } while their intersection
is given  by
 \eqn\inter{ {\underline{\rm
intersection}:}\ \ x^2 = x^1 + {v\over {\rm sin}\theta} \; x^0 =
0\ .} As already stated, the intersection is timelike, spacelike,
or null for ${v}$ respectively smaller, greater, or equal to ${\rm
sin}\theta $~.

If the intersection trajectory  is timelike,  we can bring the
point I (as well as the moving D1-brane) to rest,  by boosting
with velocity $v/{\rm sin}\theta < 1$ in the $x^1$ direction.
As the boost is in a longitudinal direction, the static brane
is left unchanged. The transformed configuration has  two static
D1-branes,  intersecting at an angle \eqn\thetaprime{
\theta^\prime= {\rm arctan}\left( {\sqrt{{\rm sin}^2\theta - v^2}
\over {\rm cos}\theta}\right) \ .} This  is a  non-supersymmetric
and unstable configuration, in which the lowest-lying state of stretched
open strings joining the two branes will be tachyonic (see for instance \bdl).

If the intersection trajectory  is spacelike,  it can be
brought to the special form $x^{2\ \prime}= x^{0\ \prime } = 0$.
This can be achieved by a Lorentz boost with velocity ${\rm
sin}\theta/v < 1$ in the $x^1$ direction. The boost again leaves 
the static brane invariant.  The transformed configuration thus
has two parallel D1-branes,  moving  with relative speed
\eqn\vprime{v^{\prime}  = {\sqrt{v^2-{\rm sin}^2\theta}\over {\rm
cos}\theta} \ , } and coinciding at $ x^{0\ \prime  } = 0$. This
is again a  non-supersymmetric configuration, with
velocity-dependent forces and an instability for pair creation of
stretched open strings \refs{\dyn,\rev }.

 The third case, that of a null worldvolume intersection,  is the one that will
interest us here. It cannot be transformed to a static
configuration, nor to one of moving but parallel D-branes.
 The two D-brane worldvolumes are in this case related
by the  `null  Lorentz transformation' \foot{We use the
conventions $J^{\mu\nu}= -i(x^\mu \partial^\nu -
x^\nu\partial^\nu)$ and $\eta^{\mu\nu}= (-++ \dots +)$.}
 \eqn\nullboost{
\Lambda(\theta)
 = {\rm exp}( -\sqrt{2}\; {\rm tan}\theta\; {\cal J})\ \ \
{\rm with}\ \ \ {\cal J} = {i\over \sqrt{2}}( J^{\;02}+ J^{12})\ .
} In the basis ${\bf x} = (x^+,\ x^2,\ x^-)$, with $x^\pm = (x^0\pm
x^1)/\sqrt{2}$\ , the generator ${\cal J}$ takes the simple form
\eqn\Jgen{ {\cal J} = \left( \matrix{ 0&0&0\cr 1&0&0\cr
0&1&0\cr}\right) \ .} The reader can easily verify that
 the
condition $ y^{2 } = 0$, with ${\bf y} = \Lambda(\theta)\;{\bf
x}$, is the same as the equation \moving\ which defines  the
embedding of the moving D1-brane, in the critical case $v= {\rm
sin}\theta$. We will show in the following section that, in contrast to  the
two other cases which are non-supersymmetric,   the null-scissors
configuration is $1/4$ supersymmetric.

A simple generalization that still preserves $1/4$ spacetime supersymmetry is
one in which the moving D1-brane is displaced in the $x^3$
direction by a distance $b$,  so that the reference brane is at
$x^3=x^4=\cdots = x^9=0$ while the moving one is at $x^3=b,\;
x^4=\cdots = x^9=0$. We will refer to this configuration as the
\lq shifted null scissors'.  The intersection point is here
replaced by the shortest linear segment joining the D-strings.
This has length $b$ and moves with speed $v_{\rm I}=1$. It is
important to stress that $b$ is the only physical parameter of
this setup. Indeed, the velocity $v ={\rm sin}\theta$ can be given
any value between 0 and 1 by  Lorentz boosting in the $x^1$
direction. We will later choose for convenience 
\eqn\conv{{\rm tan}\theta = {\pi \over \sqrt{2}}\ ,\ \ \ {\rm so \
\ that}\ \ \ \ \Lambda(\theta) = e^{- \pi \cal J}\ .}
 The boost
required to go to this special frame does not  of course  affect
the intersection worldline, which remains null.

The   `null scissors' and `shifted null scissors' configurations
are the open-string analogues  of the parabolic orbifold \hs\  and
null flux-brane \FF\ backgrounds for  closed strings. The
generator of the orbifold group in these backgrounds is the same
as  the Poincar{\'e} transformation that  relates the static and
moving branes in our setting. Part of our motivation for the
present work was to gain more insight into  the physics of the
corresponding closed-string problem. We will return to this
relationship later on.


\newsec{Duality Maps and Supersymmetry}

   The null-scissors configuration can be transformed to other
equivalent configurations by  (chains of) duality maps.
For instance, the following series of dualities:
\eqn\dualchain{ D1 \;
{\buildrel T\over \longrightarrow}
 \;\;  D5\;\; {\buildrel S\over \longrightarrow}\;\;  NS5_B\;\;
{\buildrel T\over \longrightarrow}\;\; NS5_A\;\; {\buildrel
lift\over \longrightarrow}\;\; M5\ , } maps the two D-strings of
type IIB theory to two M-theory fivebranes, which intersect  along
a five-dimensional null hypersurface. The first T-duality acts in
four transverse  dimensions, say  (3456), while the T-duality
between the type-IIB and type-IIA theories only acts on one of
these four dimensions. The final M-theory configuration  is
precisely the one considered by the authors of  reference \Acha .

 Following \Acha , let us verify that the above
configurations  leave  1/4 of the 32 supersymmetries unbroken. The
Killing spinors in the case of M5-branes  have constant asymptotic
values $\epsilon$ that must obey \eqn\killing{ {\bf \Pi} \epsilon
= {\bf \tilde\Pi}\epsilon = \epsilon\ .} Here \eqn\defpi{ {\bf
\Pi} = \gamma^{013456}\;\ \ \ \ {\rm and} \ \ \ \ {\bf \tilde\Pi}
= S(\theta)\; {\bf \Pi}\; S(\theta)^{-1}\ , } with $S(\theta)$ the
null boost \nullboost\ in the  spinor representation of
$O(1,10)$. If $S(\theta)$ were a pure boost, or a simple rotation
on a  plane, conditions \killing\ would have no solution. In our
case, using $J^{\mu\nu}= -{i\over 2}\gamma^{\mu\nu}$ and
$(\gamma^+)^2=0$,  one finds \eqn\spinorboost{ S(\theta) =
 {\rm exp}\left({{\rm tan}\theta\over \sqrt{2}} \gamma^{2+}\right)
= 1 + {{\rm tan}\theta \over \sqrt{2}}\; \gamma^{2+} .} It
follows that both conditions can be simultaneously satisfied by
spinors that obey the chiral projections
\eqn\kil{\gamma^{01}\epsilon = \gamma^{3456}\epsilon =  \epsilon\
.} These  leave  precisely 8 unbroken supersymmetries. All dual
configurations are of course
  also  1/4 supersymmetric, as are the shifted cases in which
one of the branes has been  displaced in an orthogonal direction.

 We can gain some
further  insight into the null scissors of Figure 1  by
compactifying the $x^2$ dimension   on a circle,   and then
performing a  T-duality transformation. This  inverts the value of
the radius in string units, transforms the D1-branes to D2-branes,
and maps their embedding coordinates to Wilson lines
 (see e.g. \rev).
One finds a gauge potential $A^2=0$ for the D2-brane dual to the
static string, and \eqn\tdual{  2\pi\alpha^\prime\; A^2
 = \; {v\over {\rm cos}\theta} ~x^0
+ {\rm tan}\theta\;  x^1 \    } for the D2-brane dual to the moving
one. The final configuration is illustrated in Figure 2. It has  a
pair of static parallel D2-branes, one of which carries  a
constant worldvolume electromagnetic field. For the shifted
scissors, in which one of the D-strings is displaced by a distance
$b$ in the $x^4$ direction, the dual D2-branes are also separated
by  a distance $b$ along $x^4$. Further T-dualities along
spectator directions would
 add extra dimensions to the D2-branes,
 without affecting the electromagnetic field.

\bigskip
\ifig\ebfields{A T-dual configuration of the null scissors of
Figure 1. It is obtained  by T-dualizing the $x^2$ and  $x^3$
dimensions. One of the two resulting D3-branes carries equal in
magnitude, but orthogonal electric and magnetic fields.
 The branes can be separated by a distance $b$ in
the $x^4$ dimension.} {\epsfxsize3.1in\epsfbox{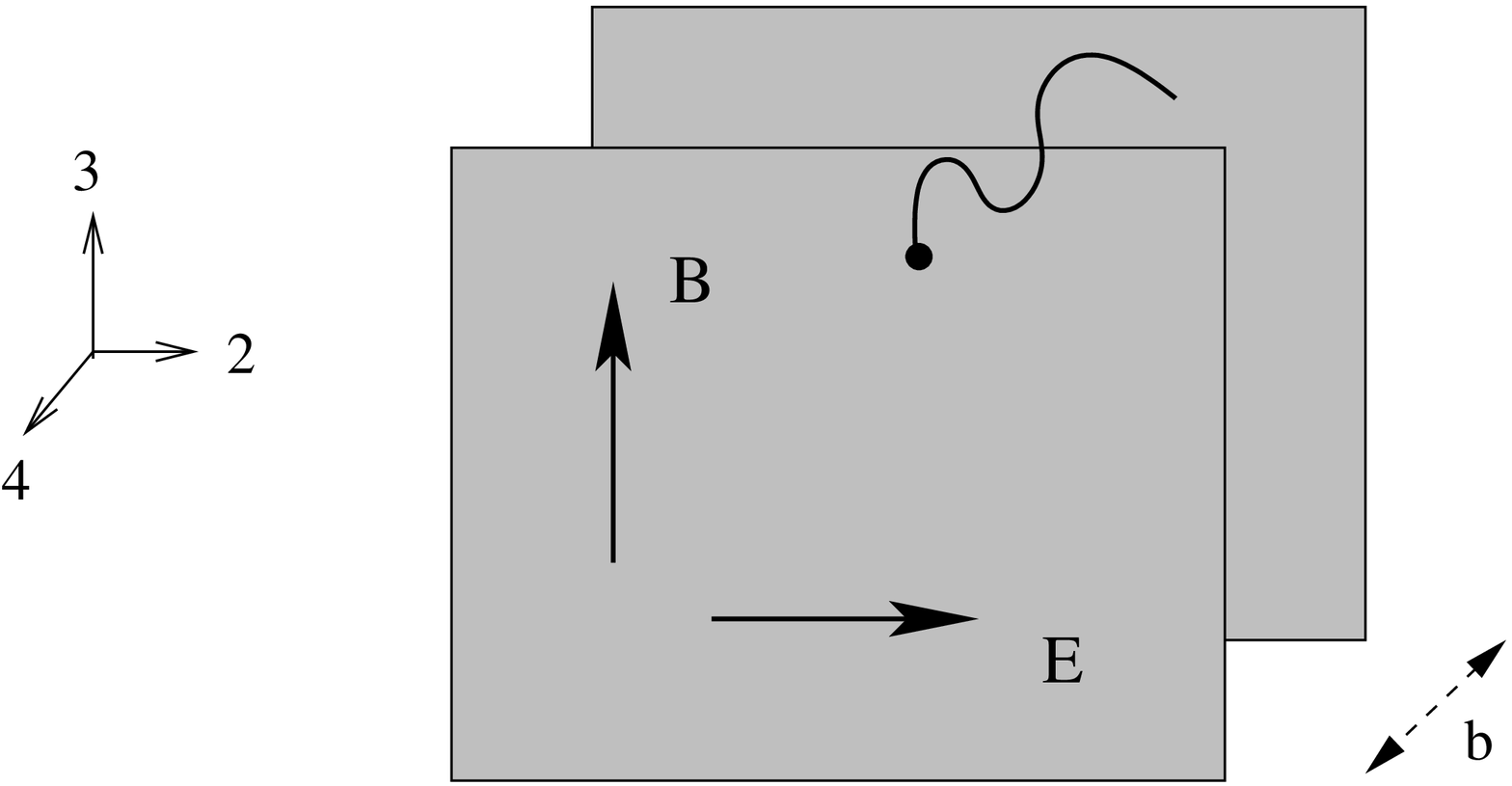}}

   The three different cases, of timelike, spacelike or null
   intersection,
can be easily seen to correspond to positive, negative or zero
values of the invariant quantity  $F_{ab} F^{ab}$.
 In the first two of these cases, a Lorentz boost can
make the field  either pure  magnetic,  or pure electric. This is
impossible in the     null case, where the field reads
\eqn\emnull{ F_{+2} =  { {\rm tan}\theta \over
\sqrt{2}\pi\alpha^\prime}  \ , \ \ \ {\rm all}\ \ {\rm other} \ \
F_{ab}=0\ . } We can think of this background as  the
infinite-wavelength limit of a plane electromagnetic wave.
 Of course,  since a Lorentz boost can rescale
$F_{+2}$,  the only invariant statement (assuming  non-compact
branes) is that the parameter ${\rm tan}\theta$ is finite and non-zero.

  We will see later, in section 6, that one can consider general
electromagnetic waves with arbitrary profiles  $F_{+2}(x^+)$~.
These will be T-dual to configurations in which the D-branes carry
transverse-displacement waves,  all travelling in the same direction
at the speed of light  
\refs{\Cal\Dab\Ma\Cho\Lun {--} \Mat} .
 Our  straight null scissors can be obtained  as 
a special limit of such  general
configurations, all of which are 1/4 supersymmetric.

 The above  T-duality transformation gives an interesting
 interpretation \dyn\  of the instability of electric
 fields in string theory \bp . An electric field is
 classically unstable in the presence of charged particles, which
 accelerate and pump energy out of the field. This is T-dual to
 the phenomenon whereby an open string joining the  two
 D-branes cannot keep pace with their intersection point if this
 moves with superluminal speed. The string will
 stretch, pumping energy out of the moving branes  and
bringing them eventually to rest. Furthermore, even the
open-string vacuum state is quantum-mechanically unstable by pair
production of such stretching  strings, a phenomenon T-dual to the
well-known Schwinger instability of electric fields.


\newsec{Open-string Spectrum}

 In this section we  analyze the open-string theory
 for the null D-brane scissors, and for its T-dual
 configurations. There are three types of open strings
in the configuration of Figure 1: those
 living on the static D-brane, those living on the moving D-brane, and
 those stretching from the static to the moving one. In the T-dual
 picture of Figure 2, the latter correspond to strings
 charged under the electromagnetic background fields, while the former two
 types of string are neutral.

  Strings with both endpoints on the same D-brane  have a standard
  spectrum,  consisting of a maximally-supersymmetric spin-1
  multiplet, plus the usual  tower of massive excitations.
  The only minor subtlety has to do with worldsheet zero
  modes. The strings on the static D-brane of Figure 1 have
 momentum $p^2=0$, \foot{Note
 that here and in what follows, $p^2 $ is the $\mu=2$ component of $p^\mu$,
and $x^2 $ is the $\mu=2$ component of $x^\mu$. The magnitude squared of the
vectors  will be written as $p_\mu p^\mu$ and $x_\mu x^\mu$.} 
while those
  on the moving brane have $p^2 = \sqrt{2}\;{\rm tan}\theta \; p^+$.
 Thus, the mass-shell condition for the latter reads
\eqn\shell{-2p^+p^- + (\sqrt{2}\;{\rm tan}\theta \; p^+)^2
+(p_\perp)^{ 2}+ M^2 = 0\ ,}
 where $p_\perp$ is the momentum in the spectator dimensions (3, \dots ,9),
along  which the D-brane  may possibly extend.
Solving condition \shell\  gives  $p^-$ in terms of $p^+$, $p_\perp$  
 and  the mass $M$. It is important here
 to realize  that   only   $p^+$ and
 $p_\perp$ are conserved momenta in the interacting theory.
 Neither $p^-$ nor $p^2$ will be  conserved in general,
 since translations of $x^+$ and $x^2$ are not symmetries of the
 D-brane background. These symmetries will be violated by open-string
diagrams whose  boundary has components on both  D-branes.

 It is interesting to see how condition \shell\ will arise in the
 T-dual picture of Figure 2. 
The 
fact that the   strings  in Figure 1   do not   wind 
in the $x^2$ direction becomes the condition in the T-dual picture 
that $p^2=0$.
In the presence of a background $F$
 field,
 open strings feel the effective metric \Abou
 \eqn\openmetric{G_{ab} = \eta_{ab} +
 (2\pi\alpha^\prime)^2 F_{ac}F_{bd}\;\eta^{cd}\ .}
 Here $a,b$ are worldvolume indices,  which are identified with a subset of
 spacetime indices in
  static gauge. Inserting \emnull\ in this expression we
 find
  \eqn\openmetr{ds^2_{\rm
open} = \eta_{ab}\; dx^adx^b+ 2\;
 {\rm tan}^2\theta \; (dx^+)^2 \ .
} It is now easy to  verify that the mass-shell condition
\eqn\masss{G_{ab}\; p^ap^b + M^2 = 0} is the same as eq. \shell ,
provided we set  $p^2=0$. The more general case,  $p^2\not= 0$,  should
be compared to strings with non-zero winding $w^2$ in the T-dual scissors.

  We turn next to the study of open strings  with one endpoint on each
of the two D-branes. These will play a crucial  role in our discussion
of stability in section 6. They are,  as will become clear,  the
open-string analogues of the twisted closed strings of references
\refs{\lmsI, \lmsII ,\Fabinger }. We 
will in fact simply adapt the analysis of these
references to our problem.

We will consider the configuration of Figure 2, and limit our
discussion  to the bosonic coordinates. Transforming back to the
null scissors via a T-duality, and extending the analysis to
worldsheet fermions, are simple matters of   detail which we
will  skip here. The boundary condition at the endpoint of an open
string coupled to a constant electromagnetic background is
\eqn\bcharge{
\partial_\sigma {X}^a =
 \pm 2\pi\alpha^\prime\; {F}^a_{\ \ b}\;  \partial_\tau {X}^b\ .  }
The sign depends on the orientation. Using equations \emnull\ and
\conv , and the definition \Jgen\ of the matrix ${\cal J}$,
 we
find the following conditions for a stretched open string:
 \eqn\bcons{
\partial_\sigma {\bf X}  = 0 \ \ {\rm at}\ \ \sigma = 0\ ,\ \ \ \
{\rm and}\ \  \ \
\partial_\sigma {\bf X} =
 \pi {\cal J}\;  \partial_\tau {\bf X}
\ \ \ {\rm at}\ \ \ \sigma = \pi\ .  }
  We  use the 3-component
vector notation ${\bf X}= (X^+, X^2, X^-)$, and we have
  dropped the seven `spectator' coordinates which obey regular
Dirichlet or Neuman conditions at both string endpoints.

The general expression for harmonic coordinates with the boundary
conditions \bcons\ can be written using
 the `spectral-flow' trick of \lmsI\ as follows:
\eqn\soln{ {\bf X}(\tau,\sigma)  = {\bf X}_{0}(\tau,\sigma) +
\sqrt{\alpha^\prime\over 2}\sum_{n\not= 0}\;
 ({\cal J}-in)^{-1} \left[\; e^{({\cal J}-in)(\tau+\sigma)} +
e^{({\cal J}-in)(\tau-\sigma)} \right]\; {\bf a}_n \ .}
 Here ${\bf a}_n$ is a triplet of oscillation amplitudes,
obeying the reality conditions ${\bf a}_{-n} = {\bf a}_{n}^*$. The
zero-mode piece is given by \eqn\zeromodes{ {\bf X}_{0}(\tau
,\sigma) = {\bf x}_0 +
 \sqrt{\alpha^\prime\over 2}\;  {\bf f}(\tau+\sigma)
+ \sqrt{\alpha^\prime\over 2}\;  {\bf f}(\tau-\sigma)\ , } with
\eqn\zeromod{ {\bf f}(y)= \int_0^y dw \; e^{{\cal J}w}{\bf a_0}\ .
} To check that \soln\ indeed satisfies the boundary conditions
\bcons\   one must use the fact that ${\cal J}$ is nilpotent. It
obeys the equation ${\cal J}^3=0$,  which implies in particular
that ${\rm tanh} (\pi {\cal J})= \pi {\cal J}$.  Using the
nilpotency of ${\cal J}$ we can  also write \zeromod\ as follows:
\eqn\zeromode{ {\bf f}(y)= y\;{\bf a_0} + {y^2\over 2}\; {\cal
J}{\bf a_0}
 + {y^3\over 6}\; {\cal J}^2{\bf a_0} \ . } The reader
can check that in the formal limit  ${\cal J}\to 0$, expression
\soln\ reduces to the standard expansion \gsw\  for Neumann
coordinates, as   expected.

 To solve for the classical motions of the open string, we need
to impose the conformal-gauge conditions. The Virasoro generators
have the usual form  in terms
 of the oscillation amplitudes:
\eqn\vira{ L_n  = {1\over 2} \sum_{m=-\infty}^\infty
 {\bf a}_{n-m}\cdot {\bf a}_m\;
\ ,  } where the dot denotes the Lorentzian inner product, and we
must here put back the spectator dimensions in the definition of
the vectors ${\bf a}_n$. Equation \vira\ follows from the fact
that $e^{x\cal J}$ is a Lorentz transformation,  so that
$$
(e^{x\cal J})^\mu_{\ \rho}\;  (e^{x\cal J})^\nu_{\ \sigma}\;
\eta^{\rho\sigma} = \eta^{\mu\nu}\ .
$$
Now since \eqn\pplus{p^+ = {1\over 2\pi\alpha^\prime}\int_0^\pi
d\sigma {\dot X}^+ = {a_0^+\over \sqrt{2\alpha^\prime}}\  } is a
conserved momentum, we can go to the light-cone gauge where $X^+ =
2\alpha^\prime p^+ \tau$. We can then solve the Virasoro
conditions,  $L_n=0$,  by  expressing the $a_n^-$ in terms of the
remaining independent amplitudes $a_n^j$.

 To gain some insight into these classical solutions, let us
assume that the string does not oscillate
in  the direction $x^2$, so that $a_n^2=0$ for $n\not=0$.
 The center of mass momenta,  defined as
in equation \pplus , are in this case:
\eqn\solnt{ {\bf p}(\tau) = {1\over \sqrt{2\alpha^\prime}}\;
\left( \matrix{ a_0^+\cr\cr
a_0^2  + a_0^+\tau  \cr\cr
 a_0^-
+ a_0^2\tau  +  a_0^+ (
{\tau^2\over 2}+{\pi^2\over 6}) \cr\cr}\right)\ .
}
If $a^+_0 = \sqrt{2\alpha^\prime} p^+$ does not vanish,   $p^2$ and $p^-$  grow
in time, as does the total energy $p^0= (p^+ + p^-)/\sqrt{2}$.
These growing modes
draw their energy from the electromagnetic background, and this solution is
only reliable until the energy grows so large that the back-reaction can no
longer be ignored (the effect of the back-reaction will be considered in
section 6).
Furthermore, a simple calculation shows that
\eqn\mass{-p^\mu p_\mu = {(\pi p^+)^2\over 3} +  M_{\perp}^2\ ,
}
where  
\eqn\massr{
M_\perp^2 = {1\over \alpha^\prime}\; \sum_{j=3}^9\sum_{n=1}^\infty\; 
 \vert a_n^j\vert^2\;  
+ \; \left({b\over 2\pi\alpha^\prime}\right)^2
}
 is  the mass
due to  transverse oscillations,  and to the shift $b$.
Equation \mass\ shows that  the total mass of the string does not grow,
and the absorbed  energy is kinetic
rather than tension energy. This is not the case
 in the equivalent  configuration of Figure 1,
where  strings with $p^+\not=0$ will
 stretch and so have both  tension and kinetic energy. As a result
 the  effective mass must grow  with time.
 The reader can check these claims  explicitly,
 by performing the T-duality
transformation  which flips
 the sign of the right-moving component of $X^2$ in
\zeromodes .

 When  $p^+=0$,  the string travels on the lightcone and it must therefore
be massless. The general solution, up to reparametrizations\foot{
If $X^+$ were not constant,  we could find a gauge such that  $X^+\propto \tau$
in a local patch. By continuity this would then reduce to
 one of the previously discussed
solutions.},
 reads
\eqn\pplusze{ X^- = \sqrt{2\alpha^\prime}\; p^- \tau\ , \ \
X^+\ \ {\rm and}\ \  X^j \ \  {\rm constant} \ . }
It describes a zero-size massless string moving in the direction 
normal to  the electric and magnetic
fields  at the speed of light. In the T-dual null scissors of Figure 1 the
string is localized at the intersection ($X^+ = X^2 = 0$). 
Clearly, such a solution does not  exist in the
shifted null-scissors configuration,   or its dual versions.
The string must stretch  in a spectator
dimension   in the shifted case, and is therefore necessarily  massive. To
summarise  our conclusions thus far:
sustainable classical string motions are only
possible in the $b=0$  case, and  they
have $p^+=p^i = M_\perp = 0$, and arbitrary  $p^-$ .
There are also classical 
growing modes, which can signal an instability, as will be discussed in
section 6.

   Consider now  the quantized  theory. Canonical commutation
 relations between $X^\mu(\sigma)$ and the conjugate variables $\Pi_\mu(\sigma)$
imply the following commutators between the oscillator
 amplitudes
and the zero modes in  the expansion \soln\  -- \zeromod :
\eqn\commt{ [ a_m^\mu , a_n^\nu] = ( m\;  \eta^{\mu\nu} - i\;
2\alpha^\prime\; F^{\mu\nu})\; \delta_{m+n,0}\ \ , \ \ \ {\rm
and}\ \ \ [x_0^\mu, a_0^\nu] = i \sqrt{2\alpha^\prime}\;
\eta^{\mu\nu}\ . } If we define $ {p}^\mu \equiv  a_0^\mu/
\sqrt{2\alpha^\prime}\; $ (these are the  classical center-of-mass
momenta at $\tau = 0$), then the algebra of zero
modes reads:  \eqn\algz{  [x_0^\mu, {p}^\nu] = i \eta^{\mu\nu}\  \ \
{\rm and}\ \ \ [ {p}^- , {p}^2] = i f \ , \ \ {\rm where}\ \ \ f
\equiv  2\alpha^\prime\; F^{2-}\;  . } This is an unfamiliar
algebra,  that we derive in the point-particle limit  in the  appendix.
We can,  in fact, easily  get rid of the `anomalous'  commutator
 by redefining   \eqn\redef{p^-\;  \Longrightarrow \;
p^- - f x^2\ .} This does not affect the remaining commutators,
but it  modifies the zero-mode contribution to the Virasoro
generator $L_0$, \eqn\modf{ p^\mu p_\mu\;\Longrightarrow\; p^\mu
p_\mu + 2 p^+ f x^2 \ ,  } and hence also the corresponding
physical-state condition.

   There are  two distinct situations~: $p^+=0$ and $p^+\not= 0$.
In the first case the wave operator \modf\ has the standard form, so that
massless states with $p^j=0$ are physical. These are the states in the
vector supermultiplet of the open superstring,  \foot{It is straightforward to
verify    that the 
tachyon subtraction and polarization conditions
are the usual ones.} travelling in the same direction as the
intersection point.  For $p^+\not= 0$,  on the other hand, 
the wave operator is that of   a particle moving in a 
potential that varies linearly with  a spatial dimension. This problem has no
normalizable eigenmodes, so the superstring will have no normalizable physical states
in this case. Physical quantum states are thus in one-to-one correspondence
with sustainable classical motions.

The mode expansion  and oscillator algebra for charged open strings are
closely related  to those  for the twisted  strings in
the parabolic orbifold and null-fluxbrane backgrounds of references 
\refs{\lmsI,\Fabinger}. 
As in these references, one can formally construct vertex operators
corresponding to a  tower of massive string states. 
However, since the corresponding  wavefunctions are not normalizable, these states
are not really asymptotic, i.e. they cannot appear as external
legs in string amplitudes.  As we will argue later, in section 6, they must
decay to massless   $p^+=0$ modes for zero shift, and they  can catalyze
a condensation of zero modes when  $b\not=0$.


\newsec{Threshold Bound States}

     We have assumed so far that the null scissors
are  made out of two infinite D-strings. In this section we will discuss
what happens when 
$(x^1, x^2)$ are coordinates on  a flat two-dimensional torus.
The angle $\theta$ and the velocity $v$ must satisfy separate
quantization conditions in this case, so the requirement that
$v={\rm sin}\theta$  imposes  a restriction on the moduli of the 
torus. As we will now show, on this special locus
of the torus moduli space the two D-strings can form a  bound state at threshold,
so that their own classical moduli space has both a Coulomb and a Higgs branch.


  To simplify the discussion, consider an orthogonal torus with radii
$R_1$ and $R_2$. Let one of the D-strings stretch in the $x^2$ direction
and move in the orthogonal direction $x^1$, while the  second string is taken
oblique and static, as illustrated on the left-hand side  of  Figure 3. If the
moving string winds $m_2$ times around the $x^2$ dimension, momentum quantization implies
that
\eqn\momq{ {2\pi m_2  R_2 T\;  v \over \sqrt{1-v^2}} = {n_1\over R_1}\ , } 
with $T$  the tension of a  D-string. Furthermore, if the  static string 
winds $l_1$ and $l_2$ times in  the $x^1$ and $x^2$ dimensions, then the  
angle $\theta$ with the $O2$ axis is given by
\eqn\angleq{ {\rm tan}\theta = {l_1 R_1\over l_2 R_2}\ .}
Combining equations \momq\ and  \angleq\ with  the
null-boost condition $\vert v\vert = \vert {\rm sin}\theta\vert $ we find 
\eqn\modulis{2\pi T R_1^2 = \left\vert {n_1l_2\over m_2 l_1} \right\vert \ .
}
This fixes one of the torus radii in terms of the quantum numbers of
the two D-strings.
Note that if we hold the 
winding numbers, radii and fundamental-string scale fixed, and take the
string coupling $g_s$ to zero, then  
 $T\sim n_1 \sim 1/g_s$. 

\bigskip
\ifig\moverot{Two D-strings related by a null boost and compactified on a torus (left).
The strings can form a bound state at threshold, represented by a single classical
D-string carrying the total winding and momentum numbers of the pair (right). 
The longitudinal momentum is carried by transverse oscillation waves.}
{\epsfxsize3.7in\epsfbox{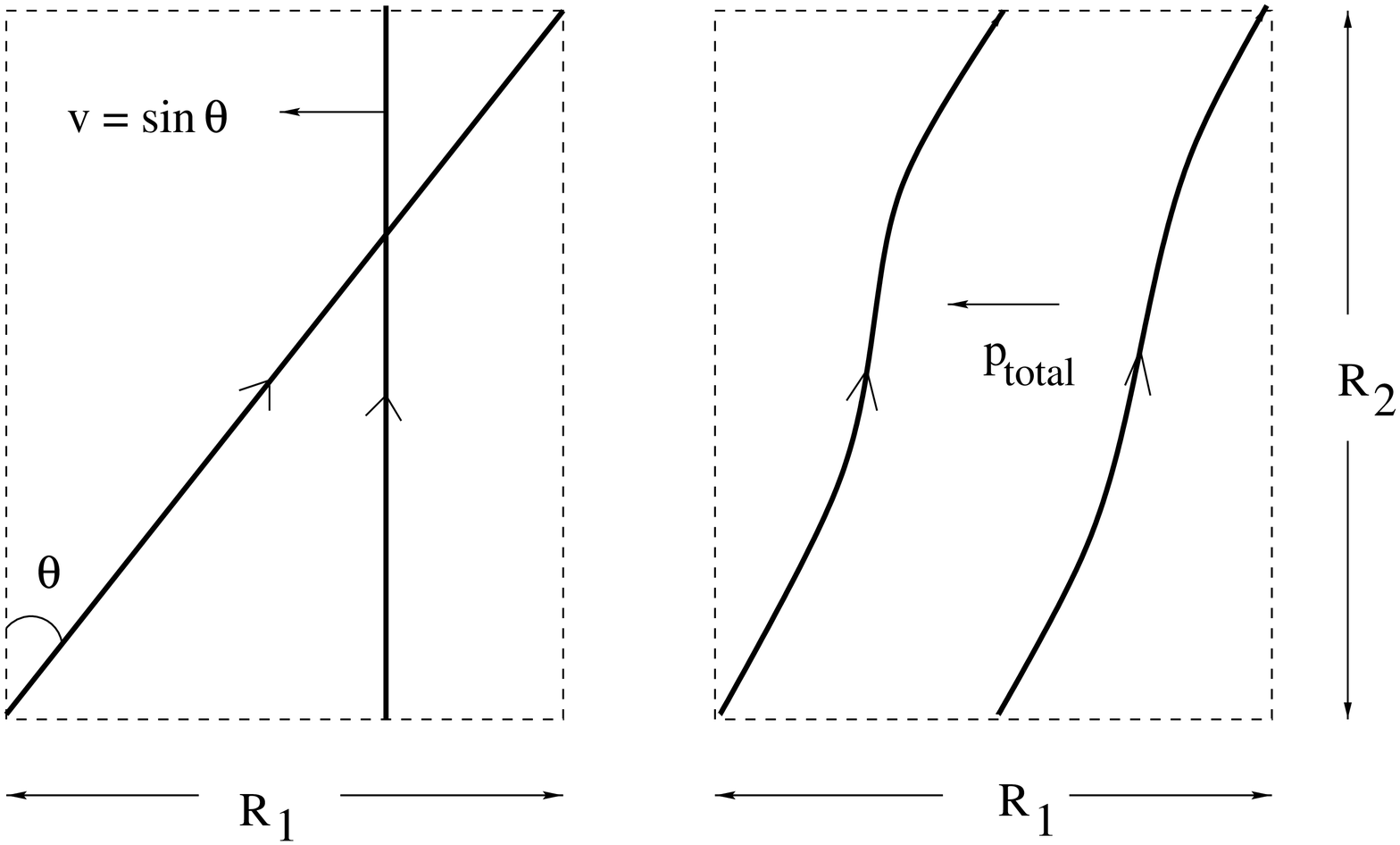}}


 Consider now a single D-string carrying the sum of the charges
of the above pair, i.e. winding $(l_1, l_2+m_2)$ times around the
two circles,  and carrying $n_1$ units of momentum in the 
direction $O1$. This  is 
illustrated,  for  the special case
 $l_1=l_2=m_2=1$, on the right-hand side  of   Figure 3. For generic  values of the
radii, the composite  string is a 
1/4 BPS subthreshold bound state of its two components. 
 We will  show that  when condition \modulis\
is satisfied, the binding energy is precisely  zero. To this end, it is convenient
to use the S-duality of the type-IIB theory, and  transform  
the D-strings to fundamental strings. The latter are  described by 
 $O(2,2)$ charge vectors 
\eqn\defch{ q \;  \equiv\; (\;\vec q_L\; ,\; \vec q_R\; ) \;
\equiv\; (\;\vec p + \vec w \; ,\; \vec p - \vec w\; ) \ , 
}
where $\vec w$ and $\vec p$ are the  winding and momentum vectors. 
In the case at hand we have  (we use here
 units such that  
$2\pi T = 1$):
$$
\underline{\rm moving}: \ \ \ \vec q_L \ \; =\; (\; {n_1\over R_1}\; 
,\;  m_2 R_2\; ) \ \ , 
\ \  \ \vec q_R\; =\;  (\; {n_1\over R_1}\; ,\;  -m_2 R_2 \;) 
$$
\eqn\chargev{
\underline{\rm static}: \ \ \  \ \vec q_L^{\ \prime}\; 
 =\; - \vec q_R^{\ \prime} \;=\; (\; l_1 R_1\;  ,\;  l_2 R_2\;  ) \ \ . 
}
The  charge vector of the bound state is equal to  the sum $q+q^\prime$, while
its mass
in the ground state is given by  
\eqn\mass{ M \; =\; {\rm max}\;  
( \;\vert \vec q_L + \vec q_L^{\ \prime}  \vert\; ,\; 
 \vert \vec q_R + \vec q_R^{\ \prime}
\vert\; )\ .
}
The binding energy is ${\cal E} = \vert \vec q_L \vert + \vert \vec q_L^{\
\prime}  \vert\ -M
= \vert \vec q_R \vert + \vert \vec q_R^{\
\prime}  \vert\ -M$
and for this 
to vanish, either  $\vec q_L^{\ \prime}$~  
must be parallel  to 
$\vec q_L$~,  or $\vec q_R^{\ \prime}$~ must be parallel to $\vec q_R$~. 
The reader will easily
verify that this condition is  equivalent to  \modulis .

We can further generalize the discussion,  so as to allow for  arbitrary
momenta to  flow  on the static and the  moving branes of Figure 3. 
Let us T-dualise in a spectator dimension to  the type-IIA theory, where 
one can recognize more readily   the   conditions for 
unbroken supersymmetry of a fundamental string,  
\eqn\susyu{ (M\gamma^0 - \vec q_L\cdot \vec\gamma)\; \epsilon_L = 0
\ \ \ \ {\rm and}\ \ \ \
(M\gamma^0 -  \vec q_R\cdot \vec\gamma)\; \epsilon_R = 0\ .
}
The states \chargev\ have $M = \vert \vec q_L\vert = \vert \vec q_R\vert $
and
$M ^{ \prime}= \vert \vec q_L^{\ \prime} \vert = \vert \vec q_R ^{\ \prime}\vert $,
 so both of the
above conditions have non-trivial  solutions. Each of these states is
 thus
1/2 BPS, while  together they only preserve 1/4 of the 32 supersymmetries. 
We have already seen this in  the dual configuration 
 of  section 3,  but let
us check  it again in the present context. 
 For strings oriented and moving
as in Figure 3, $n_1$ is a negative integer
 while  $l_1, l_2$ and $m_2$  are
positive. Using equation \modulis\  one can 
 check that the vectors $\vec
q_R$ and $\vec q_R^{\ \prime}$~ 
 point  in the same direction,
while for non-zero angle $\theta$, 
$\vec q_L$ and $\vec q_L^{\ \prime}$~ are not aligned. 
Thus, only one half of the $\epsilon_R$ Killing spinors
survive simultaneously the  projections \susyu\ for the two strings on the
left of Figure 3.  Now adding momentum 
$\vec p^{\ \prime}$  in the $-\vec w^{\ \prime}$ direction  (parallel to $\vec
w^{\ \prime}$ and in the oppoiste direction) on the
static
 string does  not break any further supersymmetries. This is because 
$\vec q_R^{\ \prime}$ will  not change  its direction, 
and   $\vert \vec q_R^{\ \prime}\vert$ is (strictly) 
 bigger than
$\vert \vec q_L^{\ \prime}\vert$. Our two strings can  thus still form a 
threshold bound state in this case. Clearly, by symmetry, 
we may also add  momentum
on the moving string. 
Notice that these  momenta on the individual strings 
point in the direction   of the intersection's motion. 
Any other small perturbation of the strings
or torus would lead to a sub-threshold bound state.

 To summarise, we have shown in  this section that
1/2 BPS   branes of type II string theory,  related by a
null boost,  can form threshold bound states when compactified on a torus. 
This remains true even  in the presence of   massless
excitations, provided these latter  propagate  in the same direction 
as the intersection 
of the branes (or
as the shortest linear segment between them, for non-zero shift).  
Though this conclusion is ultimately a  consequence of supersymmetry,
it does give a nice alternative and intuitive explanation 
for  the appearance of 1/4 BPS
threshold bound states,  at special points in the moduli space of the
toroidally-compactified type-II string theory.


\newsec{Deformations, Stability and the  Parabolic Orbifold}

 A  configuration with 8 unbroken supersymmetries, like the null 
D-brane scissors of  section 2, is normally expected to be stable.   
This is not, however, necessarily the relevant question. 
Our  D-brane configurations have a  classical moduli space, and 
one  would like to understand   whether small perturbations 
(such as open-string excitations) can push us  far 
from the original  vacuum state. 
We will now  argue that  this is 
indeed the case, and that a small perturbation can lead to a
condensation of  open strings that modifies 
the  underlying vacuum state. Put differently, though 
quantum fluctuations do not destabilize the open-string vacuum,
there can still be significant back-reaction in 
 the one- or multi-particle states.
A similar conclusion has been reached for the parabolic orbifold
in \refs{\lmsI,\Law,\lmsII,\HoPo}. The instability  discussed  here
is however different~: it is a stringy phenomenon, rather than a 
strong-gravity effect, and it a priori relevant 
for  the smooth flux-brane with $r\sim \sqrt{\alpha^\prime}$ even at
vanishingly-small string coupling.

   The basic process can be described as follows: consider a 
fundamental   open string  moving in from spatial infinity towards
the intersection (or the nearest-distance) point with both its ends
attached to  one of the 
two D-branes. Such a  string must have strictly  positive $p^+$, or else
it will be co-moving with  the nearest-distance  point. Furthermore,
 since the unbroken
supersymmetries are chiral ($\gamma^+ \epsilon = 0$)  the string is
a non-BPS excitation of the  vacuum state. Now in  the vicinity 
of the nearest-distance point, our incoming excitation can interact
and produce a pair of  open strings stretched 
between the two D-branes, which  carry  some or all of the
conserved  momentum $p^+$. The probability that this will  happen is
exponentially  small
for large shift $b$,  but it should be of order one  when 
$b\sim \sqrt{\alpha^\prime}$. 
What is the fate of the D-brane scissors  if this happens?

 Since the stretched open strings carry Chan-Paton charge, they
must  either  annihilate in pairs or else decay  to their least-excited
state.  Generically,  the strings will separate in space along
$O1$ (and possibly  along other  spectator dimensions), so 
they cannot annihilate in pairs. On the other hand, as we have seen  in
section 4, an isolated   stretched string has no normalizable states 
for non-zero shift. The string
 will therefore  accelerate and grow,  pumping energy out of
the pair of moving D-branes. The only possible outcome in the end, 
 is that back-reaction
forces the pair of D-branes to intersect, so that  the stretched open strings 
can then decay to one of their  normalizable 
$p^+=p^j=M_\perp =0$ states.

    To better understand this process, let us consider
 all marginal deformations
of the D-brane scissors which respect the 8 unbroken supersymmetries.
They correspond to massless open string states with $p^+ = p^j=0$, 
and any $p^-$. Turning on these  marginal operators on a single 
D-brane  amounts to introducing   travelling waves,  
with a profile that is an arbitrary function of $x^+$. 
Such D-branes are  known \refs{\Dab,\Lun,\Mat}
to be 1/4 BPS states of  type-II string theory -- indeed we have already
encountered their dual version in  section 5. 
To be specific,  the following generalization of the
 embeddings   \moving\  is compatible  with eight spacetime
  supersymmetries~:
$${\underline{\rm brane\  1}:}\ \ \ x^j  = Y^j(x^+) \ ,
$$
 \eqn\movingg{ {\underline{\rm brane\  2}:}\ \ \ x^j  = {\tilde Y}^j(x^+) \ ,  }
where the  $Y^{j}$ and ${\tilde Y}^j$
are arbitrary functions of the light-cone time $x^+$ (see Figure 4). 
T-dualizing   all transverse dimensions gives two  D9-branes 
carrying  electromagnetic waves,    with arbitrary
  profiles   $F_{+j}(x^+)  = dY^j/dx^+$ and 
${\tilde F}_{+j}(x^+)  = d{\tilde Y}^j/dx^+$ . 
More generally, there exist   mixed T-dual  configurations, 
in which the D-branes  carry   both transverse-displacement 
and electromagnetic 
waves, all travelling in the same direction $O1$ \refs{\Ma,\Cho,\Mat}.

\bigskip
\ifig\moverot{A deformed, twice-intersecting  null-scissors configuration,
preserving 1/4 of the spacetime supersymmetries.  
It corresponds to setting  ${\tilde Y}^j=0$ in equation \movingg\ , with a non-trivial
function ${Y}^2(x^+)$. Both  intersection points move at the speed of
light in the negative $01$ direction.
}
{\epsfxsize2.8in\epsfbox{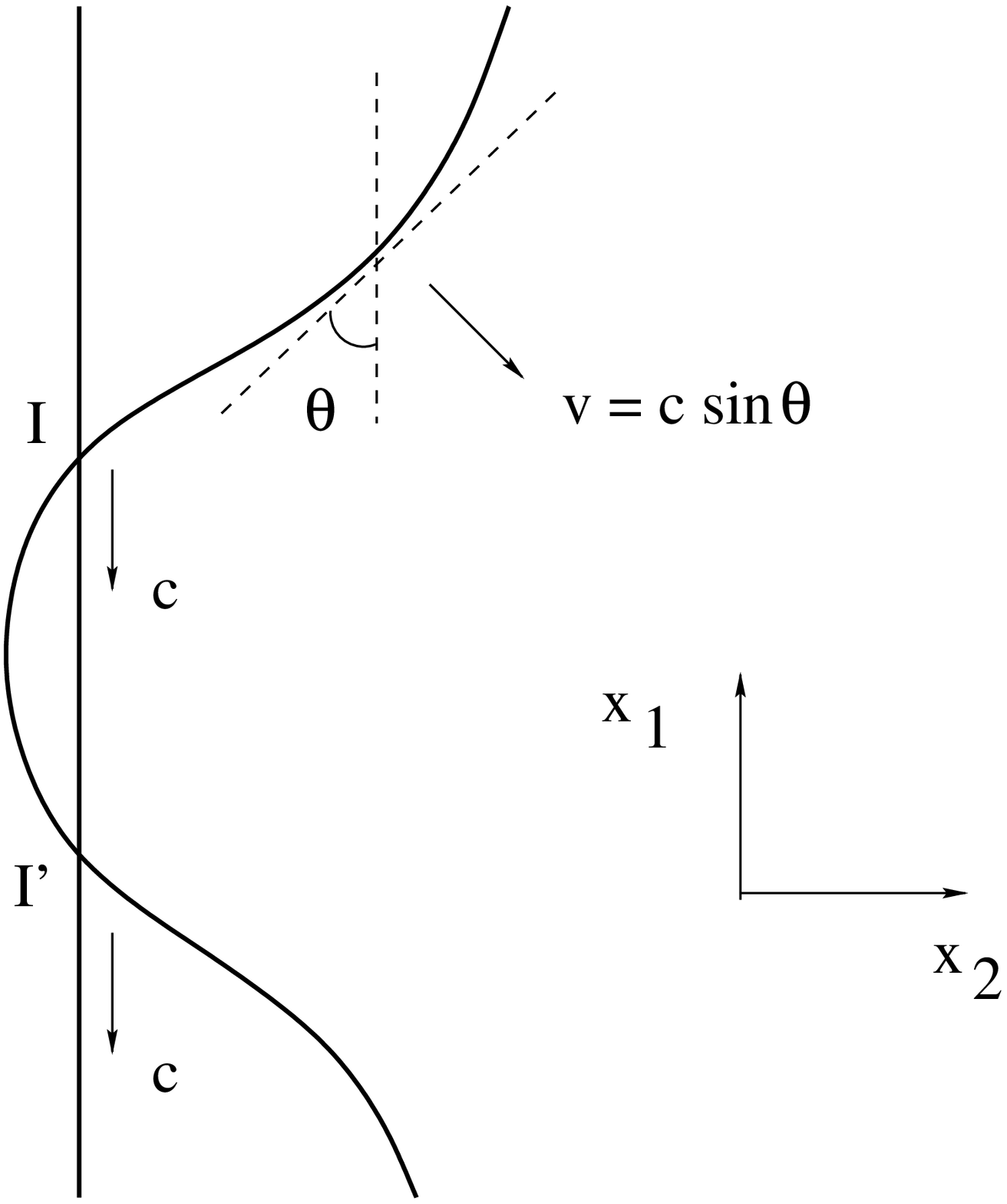}}

  That these  D-branes are good conformal boundary states follows from
the fact that the  operators  $ e^{ip^- X^+}$  
have non-singular OPEs  with each other,  as well
as with the operators $\dot X^j$. Alternatively, the Dirac-Born-Infeld
equations are  satisfied  trivially because all 
terms involving an  upper index 
 $\mu= +$  (or  a lower index $\mu = -$) 
vanish. Turning on such marginal deformations  one can
form multiply-intersecting supersymmetric `scissors', like 
the ones  illustrated  in  Figure 4.
 
  Let us go back  now to the straight scissors, on which
stretched open strings have been produced in a collision 
process.  As we  already argued, this  will force  
the two D-strings  to intersect, so that the charged
excitations can decay to  normalizable massless  states. 
What can  happen next is illustrated in Figure 5: the two D-strings
may join and split as in a  standard string-interaction process. 
The charged excitations can then be converted to 
travelling waves  of the type that was discussed above.

\bigskip
\ifig\moverot{The joining and splitting of D-strings mediated
by charged excitations. When the two D-strings touch,  the configurations
on the left and right are gauge equivalent. Such a process
converts the two relatively-boosted strings into a single long string, when
space is compactified on a torus.
}
{\epsfxsize3.2in\epsfbox{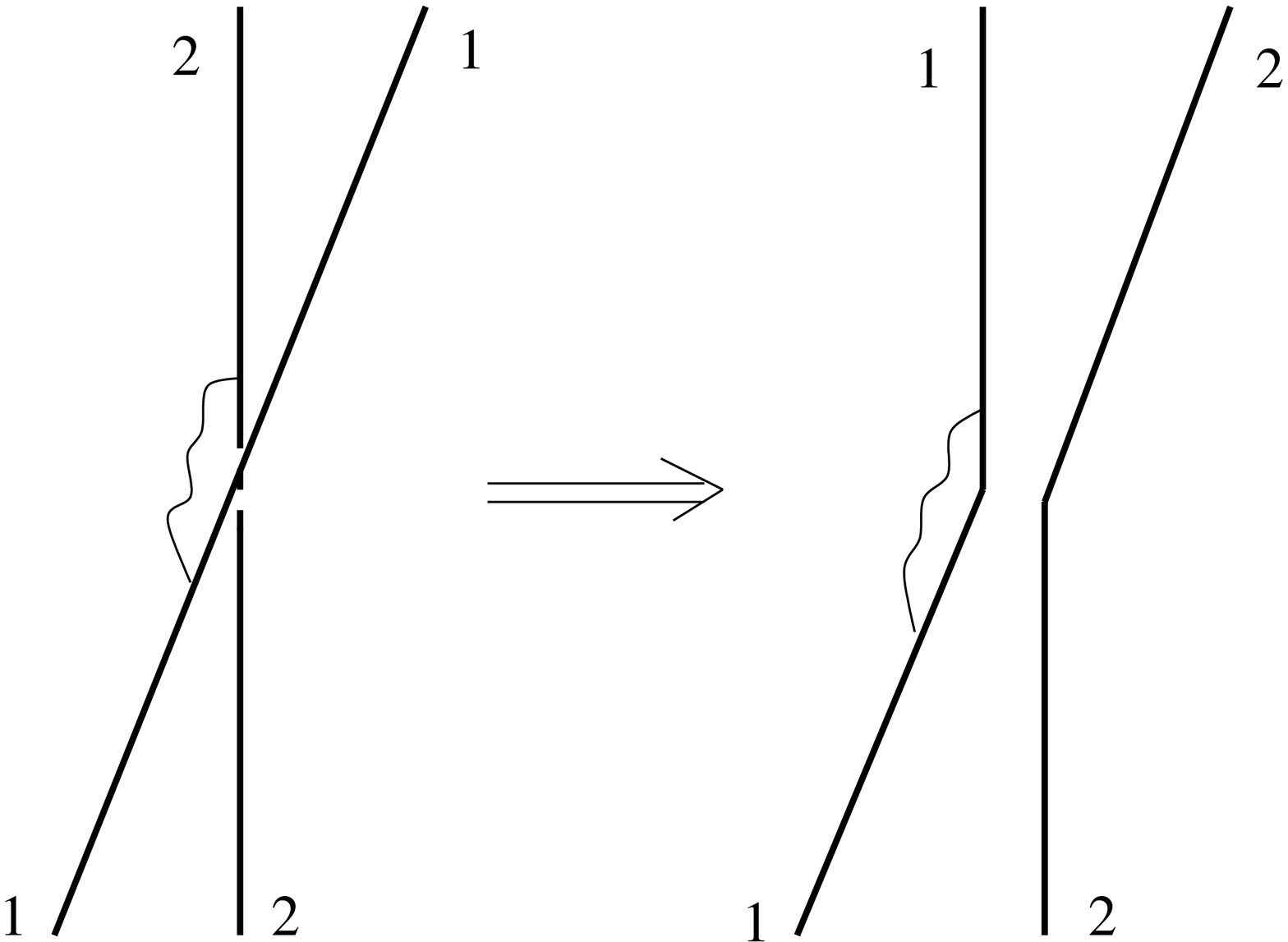}}

Note that the two configurations of Figure 5  are gauge-equivalent
when the two D-strings touch. This is most easily seen by
combining  the transverse-displacement fields  in 
$2\times 2$ matrices  ${\bf Y}^j$, which transform  in the adjoint representation
of the gauge group  $U(2)$. Then the gauge transformation
from the configuration on the the left to the one on the right
(for zero shift) can be written as follows:
\eqn\matrx{{\bf Y}^{2\ \prime}\; = \;\sqrt{2}\; {\rm tan}\theta\; 
 \left( \matrix{ \Theta(x^+)\; x^+&0\cr 0& 
\Theta(-x^+)\; x^+ \cr
}\right)
\;  =\;  g(x^+)\; {\bf Y}^2 \;  g(x^+)^{-1}\ , 
}
where $\Theta(y)$ is the Heaviside step function,
\eqn\matxx{{\bf Y}^2 \; = \;\sqrt{2}\; {\rm tan}\theta\; 
\left( \matrix{ x^+&0\cr 0&0\cr
}\right)\ , \ \ \ \ {\rm and} \ \ \ \
 g (y) = {\rm exp}\left( i\;\Theta(y)\;\sigma^1\right)
\ .}
Note also that  `wedge' D-strings,  like those on the 
right-hand side of Figure 5,  are special cases of the 
general 1/4 supersymmetric configurations \movingg\ .

  If the D-brane scissors live on a compact torus, then a non-BPS
excitation will keep colliding  with the intersection point, or
more generally with the  supersymmetric travelling waves, 
until all the momentum   $p^+$ has been carried away through
closed-string emission to  the bulk. This is the D-brane 
analogue  of the process of Hawking radiation from  
a nearly-extremal charged black hole (for a  recent review see \Dav ). 
In the process, the
 two D-strings will generically interact as in Figure 5, to form
a  single longer string for which the entropy is maximal. 
This is also consistent with our conclusions of section 5, where 
we saw that small perturbations of the D-strings, or
of the closed-string torus moduli, will force the former  to bind into 
a subthreshold bound state.

Let us now summarize our main point: one-particle excitations
can catalyze a condensation of zero modes, and modify drastically
the underlying vacuum state. The basic mechanism proceeds through the
production of stretched open strings, which force the D-branes to
intersect and recombine in a different  way. Furthermore, 
successive incoming excitations
will keep modifying the shape of the branes in the 
near-contact region.

  What does this teach us about the physics of the parabolic orbifold 
and of the  null fluxbrane? Recall that these closed-string backgrounds
are obtained by identifying points of $\IR^{9,1}$ under a Poincar{\'e}
transformation,  which  is a combination of the  null boost \nullboost\ 
and of  a translation by $2\pi r$ in a transverse dimension. 
The resulting geometry describes a circle whose radius is a function of
the light-cone time $x^+$: it starts from infinite size in the distant past,
shrinks to a minimum radius $r$  at $x^+=0$, and then re-expands
to infinite radius in the future. The intersection point of the null
scissors is replaced here by a Big-Bang singularity. Furthermore, twisted
 closed strings share  the same  properties as stretched open
strings in the  D-brane scissors~:  their   normalizable 
physical states have  $p^+ = M_\perp = 0$, and  only exist in
the unshifted case   $r=0$.

Arguing by analogy, we may expect that particles falling towards the
cosmological singularity from the asymptotic past can  produce pairs
of twisted strings,  which would then catalyze a condensation of closed-string
states. This could modify drastically the nature of the Big Bang 
singularity. If  successive infalling excitations  keep changing
the physics at  the singularity, a statistical treatment of this would
be required.
Note that this  mechanism \foot{Ideas similar to the ones advocated
 here have
been also discussed by Fabinger and McGreevy \Fabinger\ , and by
 Bachas and Nekrasov \bn\ .} differs from  the
strong-gravitational effects discussed recently in references
 \refs{\lmsI,\Law,\lmsII,\HoPo} .
It is a stringy  phenomenon  that should dominate
 for fixed $r$ and  $p^+$,   and  very weak string coupling. In this limit
there will  be no formation of black holes \HoPo\ ,  and the problem might be  
amenable to a perturbative treatment. If so, and assuming  that
the above analogy remains
good, the null D-brane scissors could be  a simple model for the resolution 
of a cosmological singularity in string theory.

\vskip 0.6cm
{\bf Acknowledgements}: We thank Jerome Gauntlett, Michael Green, 
David Mateos, 
Greg Moore and  Paul Townsend  for valuable discussions.
We are grateful to the Newton Institute, where this collaboration begun, 
for kind  hospitality  during this years' M-theory program. 
This research was partially supported by  the European Networks
`Superstring Theory' (HPRN-CT-2000-00122) and
 `The Quantum Structure of Spacetime'
(HPRN-CT-2000-00131).

\vskip 0.9cm

{\bf  Appendix~: Particle in Constant Electromagnetic Backgound}

  In this appendix we  derive   the canonical 
commutation relations for a relativistic charged
point particle coupled to a constant electromagnetic background, for
comparison with the results of section 4.
 This is described by the following
Lagrangian and constraint:
 $$
L\; =\; {1\over 2} {\dot X^\mu}{\dot X_\mu}\;  -\;
{1\over 2} F_{\mu\nu} X^\mu {\dot X^\nu}\   \ \ \ \ \
{\rm and}\ \ \ \ \ {\dot X^\mu}{\dot X_\mu} + M^2 = 0\ . 
$$
We restrict the discussion  to two space dimensions, 
so that $\mu, \nu = 0,1,2$. 
From $L$ we can derive the equations  of motion
$$
{\ddot  X}^\mu = -F^{\mu\nu}{\dot X}_\nu \ ,
$$
and the canonical commutation relations: 
$$
[ X^\mu, \; \Pi_\nu ] = i\;  \delta^\mu_{\ \nu} \ \ \ \ \
{\rm with }\ \ \ \ \  \Pi_\nu = \dot X_\nu +  
{1\over 2} F_{\nu\rho}X^\rho\ .
$$
There exist  three inequivalent possibilities, 
according to the sign of $F_{\mu\nu} F^{\mu\nu}$, which
we analyze separately.   

\vskip 0.2cm 

(i) $\underline{\rm Magnetic}$~: we can set $F_{0i}=0$~
and $F_{12}=-B$~. The 
general solution reads 
$$
X^0 = x^{ 0}+ p^0\; \tau\ , \ \ \  {X^1+iX^2\over \sqrt{2}}  = x + a\;  e^{-iB\tau} \ ,
$$  
where  the constants of the motion $x$ and $a$ are complex. 
Canonical quantization implies the following non-zero commutators:
$$
[a, a^*] = [x^*, x] = 1/B \  \ \ \  {\rm and  }\ \ \  [x^0, p^0]= -i
\ , 
$$
while  the constraint equation reads: 
$$
(p^0)^2 =  (aa^* +a^*a) B^2  + M^2\ .
$$
This leads to the usual spectrum of Landau levels, with a degeneracy proportional
to area times $B/2\pi$.

\vskip 0.2cm 
(ii) $\underline{\rm Electric}$~: we can set $F_{01}=E$~ and the remaining 
$F_{\mu\nu}=0$. The problem can be solved by analytic continuation 
from the magnetic case. The general trajectory reads:  
$$
X^2 = x^{ 2}+ p^2\; \tau\ , \ \ \ 
 {X^0\pm X^1\over \sqrt{2}}  = x^\pm + a^\pm \;  e^{\pm E\tau} \ ,
$$ 
where $x^\pm$ and $a^\pm$ are real. Canonical commutators imply
$$
[a^-, a^+] = [x^+, x^-] = -i/E \ ,  \ \ \  [x^2, p^2]= i\ ,\ \  {\rm rest = 0  }
\ . 
$$
The problem is equivalent to an inverted harmonic oscillator, so that
the constraint  
$$
   (a^+a^- +a^-a^+) E^2  + (p^2)^2 + M^2 = 0\ 
$$
does not admit any normalizable solutions.

\vskip 0.2cm 
(iii) $\underline{\rm Null}$: here we choose $F_{+2}=f$ with the remaining components
zero. The general classical solution reads
$$
X^+ = x^+ + p^+\tau\ , \ \ \ 
X^2 = x^2 + p^2 \tau +f p^+{\tau^2\over 2} \ , \ \ 
X^- = x^- + p^- \tau +f p^2{\tau^2\over 2}+ f^2 p^+{\tau^3\over 6}  \ .
$$
At time $\tau =0$ we have $X^\mu(0) = x^\mu$, and $\Pi_\mu(0) = p_\mu 
+ {1\over 2}F_{\mu\rho}x^\rho$ . Thus the canonical commutation relations 
imply: 
$$
[x^\mu, p^\nu] = i\; \eta^{\mu\nu}\ , \ \ \  [p^-, p^2]= if \ ,\ \  {\rm rest = 0  }
\ . 
$$
These agree with the relations \algz\ if we  set  $2\alpha^\prime = 1$
(so as to normalize the string's kinetic energy correctly). 
Furthermore, the mass-shell condition has the standard form,  
$$p^\mu p_\mu +M^2 = 0\ .$$
 Shifting $p^-$ as in \redef ,  maps
the problem to that of a point particle moving in a linear potential $2 p^+ f x^2$.
This   has normalizable solutions only when  
$p^+=0$~, in which case  the linear potential vanishes. 
These
normalizable  solutions describe massless  charged
particles propagating  at the speed of light in the  direction normal to
the electric and magnetic fields. 

 We can understand this phenomenon by  a  limiting procedure.
Start  with a pure magnetic field,  $F_{+2}^\prime = - F_{-2}^\prime = B/\sqrt{2}$,
and perform a boost in the  $01$ direction to a frame where
$F_{\pm 2} = e^{\pm \xi}\;  F_{\pm 2}^\prime$~. The null configuration is
 obtained in the limit  $\xi\to\infty$, $B \to 0$ with $F_{+2}\equiv f$
held finite and fixed. States with  energy $E^\prime\sim nB$
in the original problem
undergo  an infinite boost,  which sends 
 $p^+\to 0$ with $p^-$ and the energy $E$  held fixed. Transverse momentum  in
the primed frame must also be scaled  to zero like $\sqrt{B}$,  or else the 
energy in the limit  would diverge.

\footatend\vfill\supereject\immediate\closeout\rfile\writestoppt
\baselineskip=14pt\centerline{{\bf References}}\bigskip{\frenchspacing%
\parindent=20pt\escapechar=` \input refs.tmp\vfill\eject}\nonfrenchspacing \bye